\documentstyle[equations,11pt]{article}

\renewcommand{\theequation}{\arabic{equation}}
\begin{document}
\bibliographystyle{plain}
\def\m@th{\mathsurround=0pt}
\mathchardef\bracell="0365
\def\upbrall{$\m@th\bracell$}
\def\undertilde#1{\mathop{\vtop{\ialign{##\crcr
    $\hfil\displaystyle{#1}\hfil$\crcr
     \noalign
     {\kern1.5pt\nointerlineskip}
     \upbrall\crcr\noalign{\kern1pt
   }}}}\limits}
\def\theequation{\arabic{section}.\arabic{equation}}
\newcommand{\ar}{\alpha}
\newcommand{\aar}{\bar{a}}
\newcommand{\bb}{\beta}
\newcommand{\gm}{\gamma}
\newcommand{\Gm}{\Gamma}
\newcommand{\en}{\epsilon}
\newcommand{\dd}{\delta}
\newcommand{\sg}{\sigma}
\newcommand{\kp}{\kappa}
\newcommand{\ld}{\lambda}
\newcommand{\oa}{\omega}
\newcommand{\be}{\begin{equation}}
\newcommand{\ee}{\end{equation}}
\newcommand{\bea}{\begin{eqnarray}}
\newcommand{\eea}{\end{eqnarray}}
\newcommand{\bse}{\begin{subequations}}
\newcommand{\ese}{\end{subequations}}
\newcommand{\nn}{\nonumber}
\newcommand{\bR}{\bar{R}}
\newcommand{\bP}{\bar{\Phi}}
\newcommand{\bS}{\bar{S}}
\newcommand{\bU}{\bar{U}}
\newcommand{\bW}{\bar{W}}
\newcommand{\vf}{\varphi}
\newcommand{\sn}{{\rm sn}}
\begin{center}
{\Large{\bf Integrable Time-Discretisation of the
Ruijsenaars-Schneider Model$\,{}^1$}}
\vspace{.3cm}

F.W. Nijhoff\vspace{.2cm} \\
Department of Applied Mathematical Studies\\
The University of Leeds, Leeds LS2 9JT, UK\\
O. Ragnisco\vspace{.2cm} \\
Dipartimento di Fisica E. Amaldi, Universit\`a di Roma III \\
and Istituto di Fisica Nucleare, Sezione di Roma \\
P. le Aldo Moro 2, 00185 Roma, Italy\vspace{.3cm} \\
V.B. Kuznetsov \vspace{.2cm} \\
Faculteit voor Wiskunde en Informatica, Universiteit van
Amsterdam$\,{}^{2,3}$\\
Plantage Muidergracht 24, 1018 TV Amsterdam, The Netherlands
\end{center}
\footnotetext[1]{Report 94-27, Mathematical preprint series,
University of Amsterdam;  hep-th/9412170.}
\footnotetext[2]{Supported by the Nederlandse Organisatie voor Wetenschappelijk
Onderzoek (NWO).}
\footnotetext[3]{On leave from Department of Mathematical and
Computational Physics, Institute of Physics,\\
St.~Petersburg University, St.~Petersburg 198904, Russia.}
\vspace{.3cm}
\centerline{\bf Abstract}
\vspace{.2cm}
An exactly integrable symplectic correspondence is derived which in
a continuum limit leads to the equations of motion of the relativistic
generalization of the Calogero-Moser system, that was introduced
for the first time by Ruijsenaars and Schneider. For the discrete-time
model the equations of motion take the form of Bethe Ansatz equations for
the inhomogeneous spin--$\frac{1}{2}$ Heisenberg magnet.
We present a Lax pair, the symplectic structure and prove the
involutivity of the invariants. Exact solutions
are investigated in the rational and hyperbolic (trigonometric) limits
of the system that is given in terms of elliptic functions. These
solutions are connected with discrete soliton equations.
The results obtained allow us to consider the Bethe Ansatz equations
as ones giving an integrable symplectic correspondence mixing
the parameters of the quantum integrable system and the parameters of
the corresponding Bethe wavefunction.
\vskip 2cm
\pagebreak
%
%
\section{ Introduction}
\setcounter{equation}{0}
In some previous papers, \cite{Pang,Ell}, cf. also \cite{CRM},
an exact time-discretization of the famous Calogero-Moser (CM)
model, \cite{Cal}-\cite{Ols}, was introduced and investigated.
The discrete model is an integrable symplectic correspondence,
(for a definition cf. \cite{Ves}), that
in a well-defined continuum limit yields the classical equations
of motion of the CM system\footnote{It should be noted that the
discrete CM model can be inferred also from the B\"acklund
transformations for the continuous CM model, that were presented
in \cite{Woj}.  }.
A few years ago Ruijsenaars and
Schneider introduced in \cite{Ruijs1}, cf. also \cite{Ruijs2,Ruijs3},
a relativistic variant of the CM model, which is a
parameter-deformation of the original model.
The equations of motion of this system in its generic (elliptic)
form read
\bse \label{eq:RCM}
\be \label{eq:RCMa}
\ddot{q}_i = \sum_{j\ne i} \dot{q}_i \dot{q}_j v(q_i - q_j)\  \ ,
\   \ i=1,\dots,N\  ,
\ee
where the potential $v(x)$ is given by
\be\label{eq:RCMb}
v(x) = \frac{\wp^\prime(x)}{\wp(\ld) - \wp(x)}\   ,
\ee \ese
in which $\wp(x)=\wp(x|\oa_1,\oa_2)$ is the Weierstrass P-function,
$2\oa_{1,2}$ being a pair of periods.
(These are the equations of motion in the form given by Bruschi and
Calogero in \cite{BC}). This multi-particle
model is also integrable, and carries a representation of the
Poincar\'e algebra in two dimensions. Moreover, a large number
of the characteristics of the CM model are generalized in a
natural way to the relativistic case, such as the existence
of a Lax pair, a sufficient number of integrals of the motion
in involution,
and exact solution schemes in special cases.

In view of the results in
 \cite{Pang}-\cite{CRM}, a natural question
to ask is whether there exists a time-discrete version of the
model (\ref{eq:RCM}). We have to remark that on the quantum
level the transition from the usual CM system to its relativistic
counterpart, already amounts to a discretization (or $q$-deformation).
In fact, the relevant operators in the quantum relativistic model are
commuting
difference operators, rather than differential operators, cf.
\cite{Diej,Tom}. In view of the importance of quantum Calogero-Moser type of
models in the context of representation theory, and in particular
in connection with the Knizhnik-Zamolodchikov (KZ) equations
as has been revealed by a large amount of recent work,
cf. e.g. \cite{FFR}-\cite{Cher}, these difference operators yield
new interesting connections with $q$-special functions.

In this paper we will introduce a discrete-time version of the
model (\ref{eq:RCM}), in the form of an integrable symplectic
correspondence (i.e. multi-valued map) which goes in a continuum
limit to the original model. This amounts to one more
parameter-deformation of the CM model: apart from the ``spatial''
discretisation (encoded in the parameter $\ld$), the discretisation
of time constitutes another deformation where the finite step-size
in time enters as the new parameter.  The construction is based on an
Ansatz for a Lax pair, together with an elliptic version of the
Lagrange interpolation formula. We will demonstrate the integrability
of this mapping along the lines of ref. \cite{Ves,BRST}, i.e.
integrability in the sense of Liouville.
In contrast to the previous paper, \cite{Ell}, where the particle
model is related to pole solutions of discrete soliton equations, in
particular of the lattice Kadomtsev-Petviashvili (KP) equation,
the relativistic case is related to soliton solutions of
this type of discrete equations. Finally, we point out the intriguing
resemblance between the equations of motion of the discrete particle
model and Bethe Ansatz equations for integrable
spin-$\frac{1}{2}$ quantum chains of the XYZ Heisenberg model (see
\cite{Bax,FT}).
%
%
\section{Derivation of the Model}
\setcounter{equation}{0}

We will start from an {\em Ansatz} for a Lax pair of the form
\bse \label{eq:lax} \bea
L_\kp &=& \sum_{i,j=1}^N h_i h_j \Phi_\kp(q_i - q_j + \ld) e_{ij}\   ,
\label{eq:laxa} \\
M_\kp &=& \sum_{i,j} \widetilde{h}_i h_j \Phi_\kp(\widetilde{q}_i
- q_j + \ld) e_{ij}\   . \label{eq:laxb}
\eea \ese
In (\ref{eq:lax}) the $q_i$ denote the particle positions, and the
$h_i$ are auxiliary variables which we specify later. The tilde
is a shorthand notation for the discrete-time shift, i.e. for $q_i(n)=q_i$
we write $q_i(n+1)=\widetilde{q}_i$, and $q_i(n-1)=\undertilde{q_i}$.
The variable $\kp$ is the additional
spectral parameter, whereas $\ld$ is a parameter of the system.
The matrices $e_{ij}$ are the standard elementary matrices whose
entries are given by $(e_{ij})_{k\ell}=\dd_{i,k}\dd_{j,\ell}$.
The function $\Phi_\kp$ is defined as
\be \label{eq:Phi} \Phi_\kp(x)\equiv \frac{\sg(x+\kp)}{\sg(x)\sg(\kp)}\  ,
\ee
where $\sg(x)$ is the Weierstrass sigma-function, given by
\be
\sigma(x) = x \prod_{k,\ell \ne 0} (1-\frac{x}{\omega_{k\ell}})
 \exp\left[ \frac{x}{\omega_{k\ell}} + \frac{1}{2}
( \frac{x}{\omega_{k\ell}})^2\right]\ ,
\ee
with $\oa_{kl}=2k\oa_1 + 2\ell \oa_2$ and
$2\omega_{1,2}$  being a fixed pair of the primitive periods,
see e.g. \cite{Bateman}. The relations between the
Weierstrass elliptic functions are given by
\be
\zeta(x) =  \frac{\sigma^\prime(x)}{\sigma(x)}\   \  , \   \
\wp(x) = - \zeta^\prime(x)\   ,  \ee
where $\sg(x)$ and $\zeta(x)$ are odd functions and $\wp(x)$ is an
even function of its argument.
We recall also that the $\sg(x)$ is an entire function, and
$\zeta(x)$ is a meromorphic function having simple poles at
$\omega_{kl}$, both being quasi-periodic, obeying
\[ 
\zeta(x+2\omega_{1,2}) = \zeta(x) + 2\eta_{1,2}\    \ ,\    \
\sigma(x+2\omega_{1,2}) = -\sigma(x)
e^{2\eta_{1,2}(x+\omega_{1,2})}\  , 
\]
in which $\eta_{1,2}$ satisfy ~$\eta_1\omega_2 - \eta_2\omega_1
= \frac{\pi i}{2}$~, whereas $\wp(x)$ is periodic.
{}From an algebraic point of view, the most important property of
these elliptic functions is the existence of a number of functional
relations, the most fundamental being
\be \label{eq:zs}
\zeta(\ar) + \zeta(\bb) + \zeta(\gm) - \zeta(\ar +\bb + \gm)
  = \frac{  \sigma(\ar + \bb )  \sigma(\bb + \gm )
\sigma( \gm + \ar) }{ \sigma(\ar) \sigma(\bb) \sigma(\gm)
\sigma(\ar + \bb + \gm )}~ .
\ee
{}From this relation, one can basically derive all important
identities for the Weierstrass elliptic functions. For our
purpose it can be recast into the form
\be \label{eq:PHI} \Phi_\kp(x)\Phi_\kp(y) = \Phi_\kp(x+y) \left[
\zeta(\kp) + \zeta(x) + \zeta(y) - \zeta(\kp+x+y)\right]\ . \ee

The Ansatz for the Lax pair (\ref{eq:lax}) is a natural one,
in view of the fact that the matrix $L_\kp$ corresponds to the Lax
matrix of the continuum Ruijsenaars-Schneider (RS) model, cf.
\cite{BC}, whereas the matrix $M_\kp$ is a natural choice by
comparison with the earlier results obtained in \cite{Ell}, cf.
also \cite{CRM},  for the discrete elliptic CM model.

Let us now consider the compatibility of the system (\ref{eq:lax}).
Assuming the Lax equation
\be\label{eq:Lax} \widetilde{L}_\kp M_\kp = M_\kp L_\kp \  , \ee
which implies the isospectrality of the discrete flow of the
Lax matrix $L_\kp$, we have from (\ref{eq:PHI}) that
\bea
&& \sum_\ell \widetilde{h}_\ell^2 \left[ \zeta(\kp) +
\zeta(\widetilde{q}_i-\widetilde{q}_\ell+\ld)
+ \zeta(\widetilde{q}_\ell-q_j+\ld)
- \zeta(\kp+2\ld+\widetilde{q}_i - q_j)\right]   \nn \\
&& ~= \sum_\ell h_\ell^2 \left[ \zeta(\kp) +
\zeta(\widetilde{q}_i-q_\ell+\ld)
+ \zeta(q_\ell-q_j+\ld)
- \zeta(\kp+2\ld+\widetilde{q}_i - q_j)\right] \  .   \nn
\eea
Noting that the conservation law ${\rm tr}\widetilde{L} = {\rm tr}L$
implies:
\be \label{eq:conserv}
\sum_\ell \widetilde{h}_\ell^2 = \sum_\ell h_\ell^2\  , \ee
we have the identity
\be \label{eq:hqhq}
\sum_\ell\left[ \widetilde{h}_\ell^2
\zeta(\widetilde{q}_i-\widetilde{q}_\ell+\ld)
- h_\ell^2 \zeta(\widetilde{q}_i-q_\ell+\ld) \right] = -
\sum_\ell \left[ \widetilde{h}_\ell^2 \zeta(\widetilde{q}_\ell
-q_j+\ld) - h_\ell^2 \zeta(q_\ell-q_j+\ld) \right]\  ,
\ee
for all $i,j=1,\dots,N$: consequently, both sides of (\ref{eq:hqhq}) must be
independent of the (external) particle label. Thus, we find a
coupled system of equations in terms of the variables $h_i,q_i$
of the form
\bse \label{eq:qh} \bea
&&\sum_\ell\left[ \widetilde{h}_\ell^2
\zeta(q_i-\widetilde{q}_\ell-\ld)
- h_\ell^2 \zeta(q_i-q_\ell-\ld) \right] = -p\  ,
 \label{eq:qha} \\
&&\sum_\ell\left[ \undertilde{h_\ell^2}
\zeta(q_i-\undertilde{q_\ell}+\ld)
- h_\ell^2 \zeta(q_i-q_\ell+\ld) \right] = \undertilde{p}\  ,
\label{eq:qhb}
\eea\ese
where $p$ does not carry a particle label. In principle $p$ can still
depend on the discrete time-variable, but we will mostly consider it to be
constant: a different choice will be taken only in Section 6. In that case,
by eliminating  the variables $h_i$ from (\ref{eq:hqhq}), we get
a closed set of equations in terms of the $q_i$.

In order to derive this closed set of equations for the variables
$q_i$ we will make use of an elliptic version of the Lagrange
interpolation formula, that was derived in \cite{Ell}. In fact,
we have the following statement:

\paragraph{\bf Lemma:} Consider $2N$ noncoinciding complex numbers $x_l,y_l$
($l=1,\dots,N$). Then, the following formula holds true:
\bea
\prod_{l=1}^N \frac{\sigma(\xi - x_l)}{\sigma(\xi - y_l)}\,=\,
\sum_{l=1}^N \left[ \zeta(\xi - y_l) - \zeta(x - y_l)\right]
\frac{\textstyle \prod_{k=1}^N \sigma(y_l - x_k)}{\textstyle
\prod_{k=1\atop k\ne l}^N \sigma(y_l - y_k)}\  , \label{eq:Lagr} \\
{}~~~~~~ {\rm where}\ \ \sum_{l=1}^N (y_l -x_l) = 0\ ,
\nn\eea
and where $x$ stands for any one of the zeroes $x_l$ of the function
on the left-hand side.
\vspace{.5cm}

Eq. (\ref{eq:Lagr}) can be derived from an elliptic version of
the Cauchy identity for the determinants of the matrix
$(\Phi_\kp(x_i-y_j))$. This Cauchy identity was proven and used
in \cite{Ruijs3} to establish the commutativity of the quantum
integrals for the continuum RS model. Actually, this identity goes
back to Frobenius, \cite{Frob}, and can be proven also
by purely combinatoric means starting from the fundamental
identity (\ref{eq:zs}). A statement similar to the one in the lemma
was formulated in \cite{Skly1}.

What is essential in (\ref{eq:Lagr}) is the independence of the
choice of $x$, which can be easily demonstrated by using the identity
\be
\sum_{i=1}^N
\frac{\textstyle \prod_{j=1}^N \sigma(y_i - x_j)}{\textstyle
\prod_{j=1\atop j\ne i}^N \sigma(y_i - y_j)}\,=\,0\   ,
{}~~  \ \sum_{i=1}^N (y_i -x_i) = 0\ ~~~ ,
\ee
which can be found e.g. in \cite{WW}, p. 451.

Using now eq. (\ref{eq:Lagr}) and applying the lemma to the
elliptic fractional ``polynomial''
\[ \prod_{k=1}^N \frac{ \sg(\xi-q_k+\ld)\sg(\xi-\widetilde{q}_k-\ld)}{
\sg(\xi-q_k)\sg(\xi-\widetilde{q}_k)}\  , \]
 and taking $\ \xi=q_j - \ld$,  respectively $\ x=\widetilde{q}_i
+ \ld\ $,
we can by comparing with eq. (\ref{eq:hqhq}) make the identifications
\bse \label{eq:h} \bea
h_\ell^2 &=& -p\;\frac{ \prod_{k=1}^N \sg(q_\ell-q_k+\ld)
\sg(q_\ell-\widetilde{q}_k-\ld)}{\left[ \prod_{k\ne \ell}
\sg(q_\ell-q_k)\right] \prod_{k=1}^N \sg(q_\ell-\widetilde{q}_k)}\  , \\
\widetilde{h}_\ell^2 &=& p\; \frac{ \prod_{k=1}^N
\sg(\widetilde{q}_\ell-q_k+\ld) \sg(\widetilde{q}_\ell-\widetilde{q}_k-
\ld)}{\left[ \prod_{k\ne \ell}
\sg(\widetilde{q}_\ell-\widetilde{q}_k)\right] \prod_{k=1}^N
\sg(\widetilde{q}_\ell-q_k)}\  , \\
&& ~~~~~~~~~~~~~ \ell=1,\dots,N \nn
\eea\ese
from which we obtain the following system of equations
\be \label{eq:dRS}
\frac{p}{\undertilde{p}}\;
\prod_{k=1 \atop k\ne \ell}^N \frac{ \sg(q_\ell-q_k+\ld)}{ \sg(
q_\ell-q_k-\ld)} = \prod_{k=1}^N \;\frac{\sg(q_\ell-\widetilde{q}_k)
\;\sg(q_\ell-\undertilde{q_k}+\ld)}{
\sg(q_\ell-\undertilde{q_k})\;\sg(q_\ell-\widetilde{q}_k-\ld)}
\  \ ,\  \ \ell=1,\dots,N\  .
\ee
Eqs. (\ref{eq:dRS}) can be considered to be a product version of
(\ref{eq:hqhq}), and is a system of $N$ equations for $N+1$ unknowns,
$q_1,\dots,q_N$ and $p$. There is no equation for $p$ separately, and
thus it should be a priori given in order to get a closed set of
equations. The most natural choice is the one for which $p$ is
constant, i.e. independent of the discrete time-variable. For
convenience we will take it equal to the fixed value $p=-\sg(\ld)^{-1}$,
but if we are interested in a continuum limit from (\ref{eq:qh}) we should
take it of the order of the reciprocal of the discrete-time step.
As already mentioned, a dynamical choice of $p$ arises at the end of
Section 6 from the application of (\ref{eq:dRS}) to the Bethe Ansatz
equations for the XYZ Heisenberg chain.

Thus, taking $p/\undertilde{p}$ to be equal to unity in (\ref{eq:dRS}),
we obtain the equations of motion of what we would like to call the
discrete Ruijsenaars-Schneider model. It is given by a
coupled  set of algebraic equations, which, in fact, resemble
closely the Bethe Ansatz equations (BAE's) for certain integrable
quantum chains with impurities. We will make this
connection more precise in Section 6. Here, we look at
eqs. (\ref{eq:dRS}) as defining a discrete dynamical system, which
amounts to an integrable symplectic correspondence, i.e. a multi-valued
map in the sense of ref. \cite{Ves}.

Before proving the symplecticity and integrability of the
correspondence, let us show first that the interpretation of
(\ref{eq:dRS}) as a discrete version of the RS model is
justified, by showing that an appropriate continuum limit
will give us back the equations (\ref{eq:RCM}) of the continuum
model. In fact, taking
\bea
\widetilde{q}_k &\mapsto& q_k - \ld + \en \dot{q}_k + \frac{1}{2}
\en^2 \ddot{q}_k + \dots\ , \nn \\
\undertilde{q_k} &\mapsto& q_k + \ld - \en \dot{q}_k + \frac{1}{2}
\en^2 \ddot{q}_k + \dots\ , \nn
\eea
and developing with respect to the small parameter $\en$, we
obtain for  the leading term
\be \frac{\ddot{q}_\ell}{\dot{q}_\ell}=\sum_{k\ne \ell}
\dot{q}_k \left[ 2\zeta(q_\ell-q_k) - \zeta(q_\ell-q_k+\ld)
- \zeta(q_\ell-q_k-\ld)\right] \  , \ee
from which we recover eqs. (\ref{eq:RCM}) using the relation
\be \label{eq:zp}
\zeta(x+\ld) - \zeta(x) -\zeta(\ld) = \frac{1}{2}
\frac{\wp^\prime(\ld) - \wp^\prime(x)}{\wp(\ld) - \wp(x)}\  .
\ee

Let us finally remark that the non-relativistic
limit is obtained by letting the parameter $\ld$ go to zero.
In fact, in this limit the variables $h_i^2$ will behave as
\be h_i^2 \mapsto 1 + \ld \left( \sum_{k\ne \ell} \zeta(q_\ell-q_k)
-\sum_{k=1}^N \zeta(q_\ell-\widetilde{q}_k) \right) + {\cal O}(\ld^2)\ .
\label{eq:hlim} \ee
As $\ld\rightarrow 0$, we will obtain the matrix $M_\kp$ of the
non-relativistic case immediately, whereas for $L_\kp$ we thus find
\be L_\kp \mapsto \ld^{-1} + \sum_i p_i e_{ii} + \sum_{i\ne j}
\Phi_\kp(q_i-q_j) e_{ij} + {\cal O}(\ld)\  , \ee
where the $p_i$ are the expressions between brackets on the right-hand
side of the arrow in (\ref{eq:hlim}). Thus, we recover the
Lax representation of the non-relativistic case as given in
\cite{Ell}.
%
%
\section{Symplectic Structure and Integrability}
\setcounter{equation}{0}
\noindent
In order to demonstrate the integrability of the discrete-time RS model,
given by the equations of motion (\ref{eq:dRS}) together with
$\ p=\undertilde{p}\ $, we need to establish
an invariant symplectic structure. Following the philosophy of previous
papers, (cf. e.g. \cite{NPC} and references therein), this can be
assessed on the basis of an
action principle, cf. also \cite{Ves,BRST}.

It is easy to note that an action for the discrete equations
(\ref{eq:dRS}) is given by
\bse \label{eq:act}
\be
S = \sum_n {\cal L}(q(n),q(n+1))\  ,
\ee
in which
\be \label{eq:lagr}
{\cal L}(q,\widetilde{q}) =
 \sum_{k,\ell=1}^N \left[ f(q_\ell - \widetilde{q}_k)
 - f(q_\ell - \widetilde{q}_k - \ld)\right] -
 \sum_{k,\ell=1 \atop k \ne \ell }^N
 f(q_\ell - q_k + \ld)\  ,
\ee \ese
with the function $f(x)$ given by
\be \label{eq:f} f(x) = \int^x \log | \sigma(\xi) | d\xi\  .
\ee
Let us point out that the function $f$ appearing here can be considered to
be an elliptic version of the Euler dilogarithm function.
The discrete Euler-Lagrange equations
\be
\frac{\partial {\cal L}}{\partial \widetilde{q}_\ell} +
\widetilde{ \frac{\partial {\cal L}}{\partial q_\ell} } = 0\    \ ,
\    \ \ell=1,\dots ,N\  ,
\ee
are easily seen to lead to eq. (\ref{eq:dRS}). The canonical momenta $p_\ell$
are found from
\be\label{eq:mom}
\widetilde{p}_\ell = \frac{\partial {\cal L}}{\partial \widetilde{q}_\ell}
= \sum_{k=1}^{N} \left( -\log | \sg(\widetilde{q}_\ell - q_k) | +
\log | \sg(\widetilde{q}_\ell - q_k + \ld) |  \right)\  .
\ee
As a consequence we have that the symplectic form $\ \Omega=\sum_k
dp_k \wedge dq_k\ $ is invariant under the correspondence, which
implies that any branch of the correspondence defines a canonical
transformation with respect to the standard Poisson brackets given by
\be\label{eq:pb}
\{ p_k,q_\ell\} = \dd_{k,\ell}\   \ ,\   \ \{ p_k,p_\ell\} =
\{ q_k,q_\ell\} = 0\  .
\ee

We will now show that in terms of the canonical variables, the Lax matrix
$L_\kp$ in (\ref{eq:lax}) has the same form as the one of the continuum
RS model. In fact, expressing the variables $h_\ell^2$ in terms of the
canonical variables, we obtain
\be \label{eq:hp}
h_\ell^2 = e^{p_\ell} \prod_{k\ne \ell} \frac{\sg(q_\ell - q_k -\ld)}{
\sg(q_\ell-q_k)} \  ,
\ee
which we will refer to as the Ruijsenaars variables, \cite{Ruijs1}.
In terms of these
variables we have the Poisson brackets
\bea
&&\{ q_k,q_\ell\} = 0\  \ ,\  \ \{ \log h_k^2\,, q_\ell\} = \dd_{k,\ell}\  ,
\nn \\
&&\{ \log h_k^2\,,\,\log h_\ell^2\} = \zeta( q_k - q_\ell + \ld)
+ \zeta( q_k - q_\ell - \ld) - 2 \zeta(q_k - q_\ell)\,,\qquad
k\neq\ell\,   . \label{eq:qhpb}
\eea
Now, there remains not
so much work left to establish the integrability. In fact, having expressed the
Lax matrix $L_\kp$ in terms of the original variables, and having
verified that it takes the same form as in the continuous case, we
can rely on the proof of involutivity that was produced by Ruijsenaars
(in Appendix A of \cite{Ruijs1})  to assess the involutivity of the invariants
of the
discrete model as well. Naturally, the invariants of the correspondence,
given by
\be I_k = {\rm tr} L^k = \widetilde{I}_k\   \ ,\  \ k=1,\dots,N\  ,
\ee
as functions of the canonical variables are the same as in the continuum
model.

Thus, we obtain the statement of involutivity of the invariants, i.e.,
\begin{equation}
\{ I_k, I_\ell \}= \{ {\rm Tr} (L^{k}), {\rm Tr} (L^{\ell}) \} =0
\hspace{0.7cm}
{\hbox{for all}}\hspace{0.3cm} k,\ell=1,2,....\  .  \label{eq:6.29}
\end{equation}
It is a theorem by Veselov that the involutivity of the invariants
lead to the linearization of the discrete flow on tori, similar to
the continuous-time situation, cf. \cite{Ves}. (For a simple argument see also
\cite{BRST}).
%
%
\section{Exact Solutions}
\setcounter{equation}{0}

Let us now consider exact solutions of the discrete RS model in two
special cases: {\em i)} the rational limit, {\em ii)} the hyperbolic
(trigonometric) limit. Here, again, we restrict ourselves to the case
$\ p=\undertilde{p}\ $, which will then lead to an explicit integration
scheme from the Lax pair, as we shall demonstrate below. If, more
generally, we do not fix $p$, the initial value problem becomes implicit,
having to take into account initial values for the $p$ also. We shall
not pursue this latter case in this paper.

The main line of reasoning in both limiting cases is very similar
to the one followed in \cite{Pang,Ell} for the non-relativistic
discrete CM model. In the discrete situation the initial value problem is
posed by asking to find $\{q_i(n)\}$ for given
initial data  $\{q_i(0)\}$ and $\{q_i(1) \equiv \widetilde{q}_i(0)\}$,
where $q_i(n) $ denotes the position of the particles $q_i$ at the
$n$th time-step. The solution to this problem is obtained for both
the rational as well as the hyperbolic (trigonometric) limits
by solving a secular problem, namely by determining the
eigenvalues of an $N\times N$ matrix which can be explicitely
calculated from the initial data for any discrete-time value $n$.
However, one may ask the legitimate question what one has gained
from this, as one has reduced the problem of solving one set of
algebraic equations (given by the discrete equations of motion
(\ref{eq:dRS})) to another problem of solving an algebraic
equation (namely the factorization of a characteristic determinant).
In the first case, nonetheless, we have to solve the system step
by step at each iteration of the map, whence the issue of the
complexity of the solution (i.e. the growth of the number of
iterates)  becomes imminent. In view of the integrability of the
correspondence, one can at most have only polynomial growth,
as Veselov has proven in \cite{Vess}, and this is immediately
clear from the exact solution, because the only growth comes
from the permutations of eigenvalues. In fact,
there are $N!$ permutations at each iteration of the map, but
whatever be the permutation chosen  at a  certain step, it will
end into  one of the $N!$ possible permutations at the following
step: the branches  will ``cross'' each other with no divergence.

\subsection*{i) Rational Case}

This is the limit that both periods tend to infinity, i.e.
$2\oa_1\rightarrow \infty\ $, $2\oa_2\rightarrow i\infty$, in
which case we can make the substitutions
\[ \sigma(x)\rightarrow x\   \ ,\    \  \zeta(x)\rightarrow
\frac{1}{x}\  \ , \    \ \wp(x)\rightarrow \frac{1}{x^2}\  . \]
Then, the Lax matrices take the form
\be \label{eq:rlax}
L_\kp = \frac{1}{\kp} h\,^{t\!}h + L_0\    \ ,\    \
M_\kp = \frac{1}{\kp} \widetilde{h}\,^{t\!}h + M_0\  ,
\ee
where $h$ denotes the (column-)vector with entries $h_i$, the
(row-)vector $\,^{t\!}h$ being its transposed, and in which
\be \label{eq:rLM}
L_0 = \sum_{i,j=1}^N \frac{h_ih_j}{\ld + q_i - q_j }e_{ij}\    \ ,
\    \ M_0 = \sum_{i,j=1}^N \frac{\widetilde{h}_ih_j}{\ld +
\widetilde{q}_i-q_j} e_{ij}\   .
\ee
{}From the form of the Lax matrices (\ref{eq:rLM}), we can then
derive the following relations
\bse \label{eq:form}\bea
\ld M_0 + \widetilde{Q}M_0 - M_0Q &=& \widetilde{h}\,^{t\!}h\   ,
\label{eq:forma}  \\
\ld L_0 + QL_0 - L_0Q &=& h\,^{t\!}h\   ,
\label{eq:formb}
\eea\ese
where we have set
\be \label{eq:Q}
 Q=\sum_{i=1}^{N}q_i e_{ii}\  .
\ee
On the other hand, from the Lax equation (\ref{eq:Lax}) and inserting
(\ref{eq:rlax}), we obtain the relations
\bse  \label{eq:6.1} \bea
\widetilde{L}_0 M_0 &=& M_0 L_0\  ,  \label{eq:6.1a}\\
\widetilde{L}_0\widetilde{h} - M_0 h &=& -p \widetilde{h}   \  ,
\label{eq:6.1b} \\
^{t\!}hL_0 - \,^{t\!}\widetilde{h} M_0 &=& -p\,^{t\!}h \  ,
\label{eq:6.1c}  \eea \ese
together with the conservation law
$\,\widetilde{^{t\!}h\cdot h} =\,^{t\!}h\cdot h$. Consequently, we can
put
\be \label{eq:rLax}
M_0= \widetilde{U}_0 U_0^{-1}\   \  ,\   \
L_0 = U_0 \Lambda U_0^{-1}\  ,
\ee
where $U_0$ is an invertible $N \times N$ matrix, and where the matrix
$\Lambda$ is constant, $\ \widetilde{\Lambda}=\Lambda\ $,
as a consequence of (\ref{eq:6.1a}).  We assume that $\Lambda$ can be
chosen to be a diagonal matrix, but this is not essential for
what follows.
Then, introducing
\be \label{eq:Vrs}
V \equiv U_0^{-1} Q U_0\   \ ,\  \ r\equiv U_0^{-1}\cdot h\   \ ,\   \
\,^{t\!}s\equiv\,^{t\!}h\cdot U_0\    , \ee
we obtain from  (\ref{eq:6.1}) and (\ref{eq:rLax})
\be \label{eq:103}
( pI + \Lambda)\cdot\widetilde{r} = r\   \ ,\   \
\,^{t\!}s\cdot (pI + \Lambda) = \,^{t\!}\widetilde{s}\    , \ee
in which $I$ denotes the $N\times N$ unit matrix,
as well as from (\ref{eq:forma}) and (\ref{eq:formb})
\be \label{eq:0105}
\ld + \widetilde{V} - V =  \widetilde{r}\,^{t\!}s\   \ ,\   \
\ld \Lambda + [ V, \Lambda ] = r\,^{t\!}s\    ,
\ee
together with the conservation law $\ \widetilde{r\,^{t\!}s}=
r\,^{t\!}s\ $. Eliminating the dyadic $r\,^{t\!}s$ from eq.
(\ref{eq:0105}) by making use of (\ref{eq:103}), we find the
linear equation
\be
\widetilde{V} = (pI + \Lambda)^{-1} V (pI + \Lambda)\,-\,
\frac{p\ld}{pI + \Lambda}\   ,
\ee
which can be immediately solved to give
\be \label{eq:V}
V(n) = (pI + \Lambda)^{-n}\left[ V(0) - n\frac{p\ld}{pI + \Lambda}
\right] (\Lambda+pI)^n\   .
\ee
Thus, the solution of the discrete RS model in the rational case is
given by the following statement :
{\it the position coordinates $\{q_i(n)\}$ of the particles at the
discrete time $n$, evolving under the discrete equations of
motion (\ref{eq:dRS}) (in the rational limit $\sg(x)\mapsto x$),
are given by the eigenvalues of the matrix: }
\be
Q(0) - np\ld\left( pI + L_0(0) \right)^{-1}\  ,
\ee
{\it where $p$ can be chosen equal to unity without loss of
generality.}\\
The choice of $p$ only affects a scaling of the variables $h_i$
in (\ref{eq:lax}), which enter the initial value of the matrix
$L_0(0)$.

\subsection*{ii) Hyperbolic (Trigonometric) Case}

In the hyperbolic limit $2\oa_1\rightarrow \infty\ ,\ 2\oa_2=
\frac{1}{2}\pi i\, ,$
in which case we can make the substitutions
\[ \sigma(x)\rightarrow \sinh(x)\   \ ,\    \  \zeta(x)\rightarrow
\coth(x)\    \  ,\  \   \wp(x)\rightarrow \sinh^{-2}(x)\  , \]
(after an appropriate gauge transformation of the Lax matrix with a
diagonal matrix),
leading to
\be
\Phi_\kappa(x) \rightarrow \coth(\kappa) + \coth(x)\   .
\ee
In this case we obtain for the Lax matrices (\ref{eq:lax})
\be \label{L,M}
L_\kp = L_0 + (\coth \kappa + \gm ) h\,^{t\!}h\   \ ,\   \
M_\kp = M_0 + (\coth \kappa + \gm) \widetilde{h}\,^{t\!}h  \  ,
\ee
where $\gm$ can be chosen at our convenience. In this case the reduced
Lax matrices $L_0$ and $M_0$ are given by
\bse \label{eq:L0,M0} \bea
L_0 &=& \sum_{i,j=1}^N h_ih_j
\left( \coth(q_i - q_j +\ld) -\gm\right)e_{ij}\   , \\
M_0 &=& \sum_{i,j=1}^N \widetilde{h}_ih_j
\left( \coth(\widetilde{q}_i-q_j+\ld) - \gm \right) e_{ij}\   .
\eea \ese
For them, as a consequence of the `splitting off' of the terms with
the spectral parameter $\coth\kappa$, we get again the system of equations
(\ref{eq:6.1}). From eqs. (\ref{eq:L0,M0}) we derive that
\bse\label{eq:102}\bea
e^{2\ld}e^{2Q} (L_0 + \gm h\,^{t\!}h) - (L_0 + \gm h\,^{t\!}h) e^{2Q}
&=& e^{2\ld} e^{2Q} h\,^{t\!}h +  h\,^{t\!}h e^{2Q}\   , \label{eq:102a}\\
e^{2\ld}e^{2\widetilde{Q}} (M_0 + \gm \widetilde{h}\,^{t\!}h) -
(M_0 + \gm \widetilde{h}\,^{t\!}h) e^{2Q}
&=& e^{2\ld} e^{2\widetilde{Q}} \widetilde{h}\,^{t\!}h +
\widetilde{h}\,^{t\!}h e^{2Q}\   ,   \label{eq:102b}
\eea\ese
in which $Q$, $h$ and $\,^{t\!}h$ are given as before.
We again make the identifications (\ref{eq:rLax}), and
introduce $V,r,s$ as in (\ref{eq:Vrs}), for which we subsequently
derive the relations
\bse \label{eq:104}\bea
e^{2\ld}e^{2V}\left( \Lambda + (\gm-1)r\,^{t\!}s\right) &=&
\left( \Lambda + (\gm+1)r\,^{t\!}s\right) e^{2V}\  ,
\label{eq:104a}  \\
e^{2\ld}e^{2\widetilde{V}}\left( I + (\gm-1)
\widetilde{r}\,^{t\!}s\right) &=&  \left( I + (\gm+1)
\widetilde{r}\,^{t\!}s\right) e^{2V}\  .  \label{eq:104b}
\eea\ese
Choosing now $\gm=1$, and using (\ref{eq:103}) we can again eliminate
the dyadic $r\,^{t\!}s$ and explicitely solve for $\exp(2V)$. We obtain
\be e^{2\widetilde{V}}\,=\,(p + \Lambda )^{-1}e^{2V}(pe^{-2\ld} +
\Lambda )\  , \ee
which can be immediately integrated to yield
\be\label{eq:105}
e^{2V(n)} = (\Lambda + pI)^{-n} e^{2V(0)} (\Lambda + pe^{-2\ld}I)^n\  ,
\ee
the $V(0)$ and $U(0)$ as well as $\Lambda$  being determined from the
initial data $Q(0)$ and $Q(1)$. Thus, rewriting (\ref{eq:105})
we obtain the following result:
{\it the exponentials of the position coordinates $\{ e^{2q_i(n)}\}$
of the particles at the
discrete time $n$, evolving under the discrete equations of
motion (\ref{eq:dRS}) (in the  hyperbolic limit $\sg(x)\mapsto
\sinh(x)$), are given by the eigenvalues of the matrix: }
\be
(L_0(0)+pI)^{-n} e^{2Q(0)} ( L_0(0) + pe^{-2\ld}I)^n \   ,
\ee
{\it where $p$ can be chosen equal to unity without loss of
generality.}\\
The choice of $p$ only affects a scaling of the variables $h_i$
in (\ref{eq:lax}), which can be accounted for in the calculation
of the $h_i(0)$ from eq. (\ref{eq:h}).
We note again that at each discrete-time value the positions of the
particles is uniquely determined up to a permutation.

We finally remark that the trigonometric limit of the elliptic
functions $2\oa_1=\frac{1}{2}\pi$,\ $2\oa_2\rightarrow$
$i\infty$ is integrated along similar lines after doing the replacements
\[ \sigma(x)\rightarrow \sin x\    \ ,\   \ \zeta(x)\rightarrow
\cot x\   . \]
We shall omit the details.
%
%
\section{Connection with Discrete Soliton Systems}
\setcounter{equation}{0}

In \cite{Pang,CRM}, the discrete CM model was obtained from
pole-solutions of a lattice version of the Kadomtsev-Petviashvili
(KP) equation. The idea that integrable  particle models of CM
type are connected with soliton equations goes back already to the
late seventies, cf. \cite{Air} and also \cite{Chu,Krich}, but had
never been applied to
discrete soliton equations. In \cite{Ruijs1}, but maybe a bit
more transparently in \cite{BB}, a connection between the
relativistic particle model and soliton solutions of nonlinear
integrable PDE's was established. It is a natural question to ask
whether such a connection also exists on the discrete level.

In order to establish such a connection, let us investigate more
closely the trigonometric solution of the previous section.
After a gauge transformation with a diagonal matrix  of the
form $\ P=\left[ (\Lambda + pI)(\Lambda + e^{-2\ld}pI)\right]
^{1/2}\ $, we can transform $e^{2V}=P^{-n}WP^n$ into
\be \label{eq:W}
W_{ij} = -2\frac{\rho_i\rho_j}{\ld_i - \oa\ld_j}\   ,
\ee
in which $\ \rho_i\equiv (P^n\cdot r)_i=(\,^{t\!}s\cdot e^{2V}P^{-n})_i\ $,
and $\oa\equiv e^{2\ld}$. The dependence on the discrete
(time-)variable is given by
\be \label{eq:rho}
\widetilde{\rho}_i = \left( \frac{\ld_i + \oa^{-1}p}{\ld_i + p}
\right)^{1/2}\rho_i\  ,
\ee
leading to the equations for $W$
\bse \label{eq:WW} \bea
\Lambda W - \oa W \Lambda &=& -2 \rho\,^{t\!}\rho\  , \label{eq:WWa} \\
PW - \oa \widetilde{W}P  &=& -2 \widetilde{\rho}\,^{t\!}\rho\  ,
\label{eq:WWb}
\eea\ese
We can then introduce the characteristic polynomial
\be\label{eq:tau}
\tau(\xi) = \prod_{j=1}^N ( \xi + e^{2q_j}) = \det ( \xi I + W ) \  ,
\ee
and show that $\tau(\xi)$ plays the role of the tau-function
of a discrete soliton system. In fact, $W$ is the kernel of the
integral operator in the soliton sector that stands at the
basis of the so-called direct linearization approach, cf.
e.g. \cite{NCWQ}-\cite{GD}.

To derive the relevant equations directly from the  resolvent of the
matrix $W$, we proceed as follows. First, using eq. (\ref{eq:WWb}),
we can perform the following simple calculation
\bea
\frac{\widetilde{\tau}(\xi)}{\tau(\oa\xi)} &=& \det \left(
(\oa\xi I+W)^{-1} (\xi I+\widetilde{W})\right) \nn \\
&=& \det\left[ (\oa\xi I+W)^{-1} \left( \xi I + \oa^{-1}W +
2\oa^{-1}(\Lambda+pI)^{-1}\cdot\rho\,^{t\!}\rho\right) \right] \nn \\
&=& \oa^{-N} \det\left( I + (\oa\xi I + W)^{-1} (\Lambda +pI)^{-1}
\cdot\rho\,^{t\!}\rho\right)\ ,   \nn
\eea
leading to
\be \label{eq:tauv}
v(\oa\xi)\equiv 1 +\,^{t\!}\rho\cdot (\oa\xi I + W)^{-1} (\Lambda +pI)^{-1}
\cdot \rho = \oa^N \frac{\widetilde{\tau}(\xi)}{\tau(\oa\xi)}\  .
\ee
which will turn out to be one of the variables governed by the
soliton system.
In order to derive discrete soliton equations for $\tau(\xi)$ or
$v$, we introduce the $N$-component vectors
\bse \be \label{eq:u}
u_j(\xi)\equiv (\xi I + W)^{-1}\Lambda^j \cdot \rho\   \ ,\  \
\,^{t\!}u_j(\xi)\equiv\,^{t\!}\rho\cdot\Lambda^j (\xi I + W)^{-1}\   ,
\ee
as well as the scalar variables
\be \label{eq:UU}
U_{ij}(\xi)\equiv 2\,^{t\!}\rho\cdot\Lambda^j (\xi I + W)^{-1}
\Lambda^i \cdot \rho\    ,
\ee \ese
($i,j\in {\bf Z}$). Making use of the relations (\ref{eq:WW}), we can
derive the following set of recursive relations between the different
vectors $u_j(\xi)$ and  $\,^{t\!}u_j(\xi)$
\bse \label{eq:uu}\bea
\oa P\cdot u_j(\xi) &=& p\widetilde{u}_j(\oa^{-1}\xi) +
\widetilde{u}_{j+1}(\oa^{-1}\xi) + U_{j,0}(\xi)
\widetilde{u}_0(\oa^{-1}\xi)\   , \label{eq:uua} \\
\,^{t\!}\widetilde{u}_j(\xi)\cdot P &=& \oa \,^{t\!}u_{j+1}(\oa\xi)
+ p \,^{t\!}u_j(\oa\xi) - \,^{t\!}u_0(\oa\xi)
\widetilde{U}_{0,j}(\xi)\  .  \label{eq:uub}
\eea
Using the definition (\ref{eq:UU}) as well as (\ref{eq:rho}),
we can derive also
\be
p\widetilde{U}_{ij}(\xi) + \widetilde{U}_{i+1,j}(\xi)
=  p U_{ij}(\oa\xi) + \oa U_{i,j+1}(\oa\xi) -
U_{i,0}(\oa\xi) \widetilde{U}_{0,j}(\xi)\  .  \label{eq:UUU}
\ee\ese

At this point we need to reflect a moment on the role of the
discrete-time shift. The shift  $U_{ij}(\xi)\mapsto
\widetilde{U}_{ij}(\xi)$ is `labeled' by the variable $p$, which
can be identified as the (reciprocal) of the lattice parameter.
Let us assume that the variables $U_{ij}(\xi)$
do not depend on only one discrete time-variable $n$, but on
a number of them, say $n_\ar$, ($\ar=1,2,3,\dots$),
and that the corresponding discrete flows are compatible, each
associated with its own parameter $p$, i.e. $p_1,p_2,p_3,\dots\ $.
This means that $U$ is a {\em function}
of these independent time-variables, i.e. $U_{ij}(\xi)=
U_{ij}(\xi;n_1,n_2,\dots )$. In that case we have for instance that
$\widehat{\widetilde{U}}_{ij}(\xi;n_1,n_2,\dots ) = U_{ij}(\xi;n_1+1,
n_2+1,\dots )=\widetilde{\widehat{U}}_{ij}(\xi;n_1,n_2,\dots )$, etc. ,
the $~\widehat{} ~$ corresponding to the translation in the second
discrete variable. Of course, this puts extra compatibility conditions
on the eigenvalues $e^{2q_j}$ of the kernel $W$ given in (\ref{eq:W})
of the soliton solutions. In fact, the particle-coordinates $q_j$ can
be shown to obey some two-dimensional lattice equations in terms of the
shifts $\ \widetilde{}\ $ and $\ \widehat{}\ $, and the compatibility
of such equations were investigated in \cite{Ell} for the
non-relativistic situation. Here, we restrict ourselves to
obtaining from (\ref{eq:UUU})  closed-form nonlinear
partial difference equations for special elements or combinations
of elements of $U_{ij}$, i.e. partial difference versions of
the well-known soliton systems. These difference systems were
investigated in detail in a number of earlier papers, (cf.  \cite{Hans}
for a review, and references therein), on the basis of the system of
equations (\ref{eq:uua})-(\ref{eq:UUU}). Thus here
we will only present the results, referring to those papers for
their derivation.

In fact, for the special element $u_{n_1,n_2,n_3}\equiv
U_{0,0}(\oa^{-(n_1+n_2+n_3)}\xi;n_1,n_2,n_3)$, we can derive the
partial difference equation
\be\label{eq:KP}
\left( p_1 - p_2 + u_{n_1+1,n_2,n_3} - u_{n_1,n_2+1,n_3}\right)
\left( p_3 + u_{n_1+1,n_2+1,n_3}\right) \,+\,{\rm cycl.\ \ perm.}
\,=\,0\   ,
\ee
which is a lattice version of the Kadomtsev-Petviashvili (KP)
equation, cf. \cite{NCWQ,NCW}. In \cite{Ell,CRM} pole solutions of this
equation were investigated associated with a discretization of
the non-relativistic Calogero-Moser model.
The relation to the $\tau$-function $\ \tau_{n_1,n_2,n_3}\equiv
\tau( \oa^{-(n_1+n_2+n_3)}\xi;n_1,n_2,n_3)\ $ is given by
\be\label{eq:Utau}
p_1 - p_2 + u_{n_1,n_2+1,n_3} - u_{n_1+1,n_2,n_3} =
(p_1 - p_2)\frac{\tau_{n_1+1,n_2+1,n_3}
\tau_{n_1,n_2,n_3}}{\tau_{n_1+1,n_2,n_3}
\tau_{n_1,n_2+1,n_3}}\   ,
\ee
from which one can derive the bilinear equation
\be\label{eq:DAGTE}
(p_1 - p_2) \tau_{n_1+1,n_2+1,n_3}(\oa^{-1}\xi) \tau_{n_1,n_2,n_3+1}(\xi)
\,+\,{\rm cycl.\ \ perm.}\ = 0 \   ,
\ee
which was first presented in \cite{Hir}, cf. also \cite{Date}.
Eq. (\ref{eq:DAGTE}) is related also to the following version of the lattice
(modified) KP equation, \cite{Date},
\be
(p_1-p_2)\left( \frac{v_{n_1,n_2+1,n_3+1}}{v_{n_1,n_2,n_3+1}}\,-\,
\frac{v_{n_1+1,n_2+1,n_3}}{v_{n_1+1,n_2,n_3}}\right)
=
(p_1-p_3)\left( \frac{v_{n_1,n_2+1,n_3+1}}{v_{n_1,n_2+1,n_3}}\,-\,
\frac{v_{n_1+1,n_2,n_3+1}}{v_{n_1+1,n_2,n_3}}\right)
\  , \label{eq:MKP}
\ee
in terms of the variable
$\ v_{n_1,n_2,n_3}= v(\oa^{-(n_1+n_2+n_3)}\xi;
n_1,n_2,n_3)\ $, with $v$ being defined in (\ref{eq:tauv}),
(taking $p_1=p$).

Finally, we mention that special reductions of the above equations
arising by imposing additional symmetries on the soliton solutions,
will lead to the lattice soliton systems of Gel'fand-Dikii type
that were introduced in \cite{GD}. In fact, taking the parameter
$\oa$ equal to a root of unity, we can derive additional constraints
on the system of equations given by (\ref{eq:uua})-(\ref{eq:UUU}).
Thus, we find in the particular case of $\oa=-1$  for $u$ the
following lattice version of the Korteweg-de Vries equation
\be \label{eq:KdV}
(p_1 - p_2 +u_{n_1,n_2+1} - u_{n_1+1,n_2})
(p_1 + p_2 +u_{n_1,n_2} - u_{n_1+1,n_2+1}) = p_1^2 - p_2^2\  ,
\ee
cf. \cite{GD},
which is related to the following bilinear equation in terms of
the $\tau$-function
\be \label{eq:tauKdV}
(p_1+p_2) \tau_{n_1-1,n_2+1}\tau_{n_1+1,n_2} +
(p_1-p_2) \tau_{n_1-1,n_2} \tau_{n_1+1,n_2+1} =
2p_1 \tau_{n_1,n_2} \tau_{n_1,n_2+1}\  .
\ee
It is this equation that corresponds in the continuum limit exactly
to the special case considered in \cite{BB}. Similar equations can
be derived for other values of $\oa$ when $|\oa |=1$, starting from
the results presented in \cite{GD}.
%
%
\section{Connection with Bethe Ansatz Equations}
\setcounter{equation}{0}

We already remarked above that the equations of motion of the
discrete Ruijsenaars-Schneider model, (\ref{eq:dRS}), resemble
closely the form of Bethe Ansatz equations (BAE's) for certain
integrable quantum models. In this section we will make this
connection more precise.
Let us first focuss on the hyperbolic limit of eqs. (\ref{eq:dRS}),
and connect it to the BAE's for the XXZ
spin--$\frac{1}{2}$ Heisenberg magnet, cf. e.g. \cite{Bax}-\cite{Bog}
and references therein. After
that, we will show that the general form of (\ref{eq:dRS}) in the
elliptic case is connected to the generalized Bethe Ansatz (BA)
for the Heisenberg XYZ model, proposed first in \cite{FT} and
developed further in \cite{Takebe}. We will not treat the correspondence
in the rational (or XXX) case since it can be obtained by a simple limit from
the hyperbolic case.

Let us, thus, consider the quadratic $R$-matrix  algebra
\be \label{eq:RTT}
R^{(12)}(u-v) T^{(1)}(u)T^{(2)}(v) = T^{(2)}(v)T^{(1)}(u)
R^{(12)}(u-v)\  ,
\ee
which is one of the central relations in the quantum inverse
scattering method (QISM), (cf. \cite{Fadd} for an early, and \cite{Bog,F}
for more recent reviews), and let us consider
the $R$-matrix for the spin-$\frac{1}{2}$ XXZ Heisenberg magnet,
\be\label{eq:R}
R^{(12)}(u) = \left( \begin{array}{cccc}
a&0&0&0\\ 0&b&c&0\\ 0&c&b&0\\ 0&0&0&a
\end{array} \right) \  ,
\ee
in which
\[ a=\sinh(u+\eta)\  \ ,\  \ b=\sinh u \  \ ,\  \
c=\sinh \eta\   .  \]
The quantum $L$-operator on each site $k$ of the spin chain has
the form
\be\label{eq:L}
L_k(u) = \left( \begin{array}{cc}
\sinh( u + \frac{1}{2}\eta \sg_k^3) & \sinh(\eta) \sg_k^- \\
\sinh(\eta) \sg_k^+ & \sinh( u - \frac{1}{2}\eta \sg_k^3)
\end{array} \right) \  ,
\ee
where $\ \sg_k^\pm = \frac{1}{2}(\sg_k^1 \pm i\sg_k^2 )\ $,
$\sg_k^{1,2,3}$ being the Pauli matrices on site $k$, (i.e.
$\sg^\ar_k={\bf 1}\otimes \dots\otimes \sg^\ar\otimes \dots
\otimes {\bf 1}$ with $\sg^\ar$ on the $k$th entry of the
tensor product). Let us now construct the following monodromy
matrix
\be\label{eq:T}
T(u)= L_M(u-\dd_M) \dots L_2(u-\dd_2) L_1(u-\dd_1)=
\left( \begin{array}{cc} A(u) & B(u)\\ C(u) & D(u)
\end{array} \right) \  ,
\ee
where $M$ is the length of the spin chain, and in which the
$\dd_k, k=1,\dots,M$ are impurity parameters.
Both the $L_k(u)$ as well as $T(u)$ obey the relation
(\ref{eq:RTT}).
Let us now recall the standard algebraic Bethe Ansatz construction,
\cite{Fadd}-\cite{F}, i.e. there is a vacuum
state $\Omega_M$, which is an eigenvector of the diagonal entries
$A(u),D(u)$ of the monodromy matrix, and which is annihilated
by the operator $C(u)$
\be \label{eq:ADC}
A(u) \Omega_M = a(u) \Omega_M\   \ ,\   \
D(u) \Omega_M = d(u) \Omega_M\   \ ,\   \
C(u) \Omega_M = 0\  .
\ee
As the monodromy matrix is a comultiplication of the $L$-operators
along the sites of the chain, we have that
\be
a(u) = \prod_{k=1}^M a_k(u)\   \ ,\   \ d(u) = \prod_{k=1}^M
d_k(u)\  ,
\ee
where the functions $a_k(u)$ and $d_k(u)$ are the eigenvalues
of the vacuum for the diagonal entries of the corresponding
$L$-operator on site $k$. The algebraic Bethe Ansatz amounts to the
creation of an eigenstate of the form
\be \label{eq:Psi}
\Psi(q_1,\dots , q_N) = \prod_{j=1}^N B(q_j) \Omega_M
\ee
of the trace of the monodromy matrix $A(u) + D(u)$,
\be\label{eq:trace}
\left( A(u) + D(u)\right) \Psi = t(u) \Psi\  ,
\ee
with $t(u)$ being the corresponding eigenvalue.
We have then the following proposition:
\paragraph{Proposition:} $\Psi$ is an eigenfunction of the trace
of the monodromy matrix iff the numbers $q_j$ satisfy the
Bethe Ansatz equations
\be \label{eq:BAE}
\prod_{k=1 \atop k\ne \ell}^N
\frac{ \sinh(q_\ell-q_k+\eta)}{ \sinh( q_\ell-q_k-\eta)} =
\frac{a(q_\ell)}{d(q_\ell)}\,,\qquad \ell=1,\dots , N\  .
\ee
\vspace{.3cm}

In the case of the monodromy matrix (\ref{eq:T}), the functions
$a(u)$, $d(u)$ take the form
\be \label{eq:ad}
a(u) = \prod_{k=1}^M \sinh(u-\dd_k+\frac{1}{2}\eta)\   \ ,\   \
d(u) = \prod_{k=1}^M \sinh(u-\dd_k-\frac{1}{2}\eta)\   ,
\ee
which in the special case that $M=2N$ lead to the BAE's
\be \label{eq:BAEs}
\prod_{k=1 \atop k\ne \ell}^N \;
\frac{ \sinh(q_\ell-q_k+\ld)}{ \sinh( q_\ell-q_k-\ld)} =
\prod_{k=1}^N \;\frac{\sinh(q_\ell-\widetilde{q}_k)\;
\sinh(q_\ell-\undertilde{q_k}+\ld)}{
\sinh(q_\ell-\undertilde{q_k})\; \sinh(q_\ell-\widetilde{q}_k-\ld)
}\  \ ,\  \ \ell=1,\dots,N\  ,
\ee
i.e. eq. (\ref{eq:dRS}) in the hyperbolic limit ($\ \sg(x)\mapsto
\sinh(x)\ $) with the identifications $\ld=\eta$, $\widetilde{q}_k=
\dd_k-\frac{1}{2}\eta$, $\undertilde{q_k}= \dd_{k+N}+\frac{1}{2}\eta$,
($k=1,\dots,N$). Thus, the discrete RS system in the hyperbolic limit
can be reinterpreted
as the BAE's for the spin-$\frac{1}{2}$ XXZ Heisenberg chain with
the number of spins (impurities) equal to twice the excitation number of the
eigenstate. We should stress that within such an identification the
impurities play the role of the $N$ particle coordinates at times $n-1$
and $n+1$ ($\undertilde{q}$ resp. $\widetilde{q}$), which means that the
integrable correspondence mixes the parameters of the quantum model
(i.e. the impurities) and the parameters of a solution given by the Bethe
wavefunction.

To make the connection with the elliptic case we have to consider
the QISM for the XYZ Heisenberg model that was treated in \cite{FT}
for the spin--$\frac{1}{2}$ situation. For arbitrary spin one needs to
consider the algebras introduced by Sklyanin in \cite{Skly}. However,
we are interested here in the inhomogeneous spin--$\frac{1}{2}$ chain,
i.e. including arbitrary impurities. Both extensions were treated
in a recent paper by Takebe \cite{Takebe}. It was shown that the
generalized Bethe Ansatz of Takhtajan and Faddeev is applicable to
this model. The corresponding $R$-matrix is given by
\be\label{eq:RR}
R^{(12)}(u) = \left( \begin{array}{cccc}
a&0&0&d\\ 0&b&c&0\\ 0&c&b&0\\ d&0&0&a
\end{array} \right) \  ,
\ee
in which
\[ a=\sn(u+2\eta;k)\  \ ,\  \ b=\sn(u;k)\  \ ,\  \
c=\sn(2\eta;k)\  \ , \  \ d= k \sn(2\eta;k) \sn(u;k) \sn(u+2\eta;k)\   . \]
The quantum $L$-operator on each site $k$ of the spin chain now has
the form
\be\label{eq:LL}
L_k(u) = \left( \begin{array}{cc}
w_0 {\bf 1} + w_3\sg_k^3 & w_1\sg_k^1 - i w_2 \sg_k^2 \\
w_1\sg_k^1 + iw_2\sg_k^2 & w_0{\bf 1} - w_3\sg_k^3
\end{array} \right) \  ,
\ee
where
\[ w_0+w_3=\sn(u+\eta;k)\  \ ,
\  \ w_0 - w_3=\sn(u-\eta;k)\   , \]
\[ w_1+w_2=\sn(2\eta;k)\  \ , \  \ w_1-w_2= k \sn(2\eta;k)
\sn(u+\eta;k) \sn(u-\eta;k)\   . \]
The monodromy matrix takes again the form (\ref{eq:T}).
The generalization of the algebraic Bethe Ansatz to the
model described in (\ref{eq:RR}), (\ref{eq:LL}) is known only for the
special eigenfunctions {\em for which the number of excitations is half
the number of spins in the chain}. Of course, this constitutes only
a special class within the total set of eigenfunctions, but no further
results exist to date. Curiously, this is precisely the case in which
we have a connection with the discrete RS model!

To be more precise, the generalized BA, developed in \cite{FT,Takebe},
consists of
constructing eigenfunctions of the trace of the monodromy matrix
$\ A(u)+D(u)\ $ after performing a gauge transformation of the form
\be \label{eq:gauge}
T(u) \mapsto T_{m,m^\prime}(u) = S_m(u)^{-1}T(u) S_{m^\prime}(u) =:
\left( \begin{array}{cc} A_{m,m^\prime}(u) & B_{m,m^\prime}(u)\\
C_{m,m^\prime}(u) & D_{m,m^\prime}(u)
\end{array} \right) \  ,
\ee
where $S_m(u)$ is an appropriate scalar matrix. (Note that we deviate from
the notations in \cite{FT,Takebe} to be consistent with
the notations used in earlier
sections of the present paper.) Instead of a single
vacuum there is now a set of generating (vacuum) vectors $\Omega_{M,n}$
for which one has
\be \label{eq:vacc}
A_{M,n}(u) \Omega_{M,n} = a(u) \Omega_{M,n-1} \   \ ,\   \
D_{M,n}(u) \Omega_{M,n} = d(u) \Omega_{M,n+1} \   \ ,\   \
C_{M,n}(u) \Omega_{M,n} = 0\  .\ee
The generalized Bethe eigenfunctions are linear combinations of the
vectors
\be \label{eq:Psii}
\Psi_n(q_1,\dots , q_N) = B_{n+1,n-1}(q_1) \dots B_{n+N,n-N}(q_N)
\Omega_{M,n-N}\    ,
\ee
of the form
\be \label{eq:eigen}
\Psi_\theta(q_1, \dots ,q_N) = \sum_{n=-\infty}^\infty e^{2\pi i n\theta}
\Psi_n(q_1,\dots ,q_N)\   ,
\ee
where $\theta$ is a free parameter. As it was pointed out in \cite{FT},
it is possible that the series (\ref{eq:eigen}) is summable to zero
for all $\theta$ except for a finite number of values $\theta_j$.
(The results of Baxter \cite{Bax} show that among the $\theta_j$
there is also the value $\theta=0$.) Then for such values of $\theta$
one can produce the corresponding eigenfunctions, provided
that the excitation numbers $q_1,\dots,q_N$ obey the following BAE's (cf.
\cite{Takebe})
\be \label{eq:BAEss}
X\;\prod_{k=1 \atop k\ne \ell}^N \;
\frac{ \sg(q_\ell-q_k+2\eta)}{ \sg( q_\ell-q_k-2\eta)} =
\prod_{k=1}^{2N} \;\frac{\sg(q_\ell-\delta_k+2\eta)}
{\sg(q_\ell-\delta_k)}\  \ ,\  \quad \ell=1,\dots,N\  ,
\ee
\be \label{more1}
X={\rm exp}\left[-4\pi i\theta+
2\eta\frac{\eta_2}{\omega_2}\left(\,\sum_{k=1}^N(2q_k
+2\eta)-\sum_{k=1}^{2N}\delta_k\right)\right]
\ee
in the special case\footnote{
The relation $M=2N$ follows because of spin-$\frac{1}{2}$
situation, while becoming $N=\ell M$ for the arbitrary spin $\ell$
(cf. \cite{Takebe}).} that $M=2N$, and
re-expressing the equations derived in \cite{Takebe} in terms of
$\sigma$-function. Eqs. (\ref{eq:BAEss}) correspond
to the equations (\ref{eq:dRS})
for the discrete RS model in the elliptic case by the
following identifications: $\ld=2\eta$, $\undertilde{q_k}=
\dd_k$, $\;{\widetilde{q}}_k= \dd_{k+N}-2\eta$,
($k=1,\dots,N$), together with
\be \label{more2}
\frac{p}{\undertilde{p}} = {\rm exp}\left[-4\pi i\theta -
\lambda\frac{\eta_2}{\omega_2}\left({\undertilde B}+{\widetilde B}
-2B\right)\right],\qquad B=\sum_{k=1}^N q_k\  .
\ee
The sum of the particle positions $B$ plays the role of the boost
generator of the underlying Poincar\'e algebra, cf. \cite{Ruijs1}.
So, the generalized Bethe Ansatz equations correspond to the discrete-time
system with the dynamics of $p$ given in terms of the dynamics of
$q_k$'s. The results of the Section 3 can be generalized to incorporate
this case. In fact, we need to add now the term
\be
4\pi i\theta B - \frac{\eta_2}{2\oa_2}\ld (\widetilde{B}-B)^2   \  ,
\ee
to the Lagrangian (\ref{eq:lagr}) in order to get the correct factor from
the Euler-Lagrange equations. If we now calculate the canonical momenta
from (\ref{eq:lagr}) we get an extra term
$\ -\frac{\eta_2}{\oa_2}\ld (\widetilde{B} - B)\ $
on the right-hand side in (\ref{eq:mom}), but this factor cancels again
when we express the $h_\ell^2$ in terms of the canonical momenta,
because of the $p$ entering into the eqs. (\ref{eq:h}).
Thus, the final result (\ref{eq:hp})
remains unaltered\footnote{This effect can be viewed as a consequence
of the relativistic invariance of the model.}. Consequently, we still
have the same integrals
in terms of the canonical variables, and
the Ruijsenaars variables do not change, hence they are in involution and we
have the integrable correspondence.

It is amazing that exactly in the restricted case where the generalized
BA is known to apply, i.e. the case that the number of impurities
is equal to twice the number of excitations, we have a connection with
an integrable correspondence of RS type. This might also suggest that
there exists a deeper relation between integrable quantum models
solvable by the QISM on the one hand and classical integrable
multiparticle models on the other hand. The connection provided here
might yield an explanation for the existence conditions of the
generalized BA of Faddeev and Takhtajan and of Baxter's solution of the
eight-vertex model.
%
%
\section{Discussion}

The connection exhibited in the previous section between the BAE's for the
XXZ and XYZ model
with impurities and the discrete relativistic multi-particle model
is very remarkable. A similar connection seems to exist also in the
non-relativistic case studied in \cite{Pang,Ell}, which yields a
link between the equations of motion of the discrete CM model and BAE's
for the Gaudin model. In both cases, we
thus obtain a new interpretation of the BAE's for a quantum solvable
model, namely as a classical integrable dynamical model with discrete time.
Such interpretations, together with the exact solutions of the discrete
equations of motion that were derived in section 4, might lead to
possible ways of `solving' the BAE's, (which in principle constitutes a set
of transcendental equations), without having to perform some
thermodynamic limit, cf. \cite{Bog}. This,
however, requires a reformulation of the initial value problem
for the discrete model rather as a boundary value problem
on the discrete-time chain. Although it is not clear at this stage
that such boundary value problems are explicitely solvable as it was
the case in Section 4, we nevertheless
conjecture that this intriguing dynamical interpretation of the
Bethe Ansatz equations will shed new light on the solvability of the
corresponding quantum models. Furthermore, it would be interesting
to study generalizations of our results, for instance in the directions
of finding discrete particle models associated with higher-spin
models and higher-rank associated Lie algebras (nested Bethe Ansatz),
as well as investigating in this light the quadratic algebras associated
with boundary conditions for quantum integrable systems (reflection
equation algebras), cf. \cite{Skly2}. The latter might lead to
time-discretizations of $BC_N$ type of RS models, which were
investigated in \cite{Diej,Tom}.

\noindent
\subsection*{Acknowledgement}

FWN would like to thank the members of the Dipartimento di F\`isica
E. Amaldi of the Universit\`a di Roma III for their hospitality
during a visit where this work was completed. He would also like to
thank the Alexander von Humboldt Foundation for financial
support at an earlier stage of the work, and especially Prof. B.
Fuchssteiner for his hospitality during his stay at the University of
Paderborn. Furthermore, he would like to
express his great indebtness to Dr. G.D. Pang, as this work is a
natural sequel to results obtained in collaboration with him.
\pagebreak

\end{document}